# Identifying the genetic basis of antigenic change in influenza A(H1N1)

William T. Harvey[a], Donald J. Benton[b,1], Victoria Gregory[b,1], James P. J. Hall[c,2], Rodney S. Daniels[b,1], Trevor Bedford[d], Daniel T. Haydon[a], Alan J. Hay[b,1], John W. McCauley[b,1] and Richard Reeve[a,e,*]

[a]Boyd Orr Centre for Population and Ecosystem Health and Institute of Biodiversity, Animal Health and Comparative Medicine, College of Medical, Veterinary and Life Sciences, University of Glasgow, Glasgow G12 8QQ, UK; [b]The Crick Worldwide Influenza Centre, The Francis Crick Institute, Mill Hill Laboratory, The Ridgeway, Mill Hill, London NW7 1AA, UK; [c]Institute of Infection, Immunity and Inflammation, College of Medical, Veterinary and Life Sciences, University of Glasgow, Glasgow G12 8QQ, UK; [d]Vaccine and Infectious Disease Division, Fred Hutchinson Cancer Research Center, Seattle WA 98109, USA; [e]The Pirbright Institute, Pirbright, Woking, Surrey GU24 0NF, UK.

[1] formerly WHO Collaborating Centre for Reference and Research on Influenza, Division of Virology, MRC National Institute for Medical Research, The Ridgeway, Mill Hill, London NW7 1AA,UK.
[2] Present address: Department of Biology, University of York, York, UK.
[*] To whom correspondence should be addressed:
E-mail: richard.reeve@glasgow.ac.uk
Address: Graham Kerr Building, University of Glasgow, Glasgow, G12 8QQ, UK.

## Abstract

Determining phenotype from genetic data is a fundamental challenge. Influenza A viruses undergo rapid antigenic drift and identification of emerging antigenic variants is critical to the vaccine selection process. Using former seasonal influenza A(H1N1) viruses, hemagglutinin sequence and corresponding antigenic data were analyzed in combination with 3-D structural information. We attributed variation in hemagglutination inhibition to individual amino acid substitutions and quantified their antigenic impact, validating a subset experimentally using reverse genetics. Substitutions identified as low-impact were shown to be a critical component of influenza antigenic evolution and by including these, as well as the high-impact substitutions often focused on, the accuracy of predicting antigenic phenotypes of emerging viruses from genotype was doubled. The ability to quantify the phenotypic impact of specific amino acid substitutions should help refine techniques that predict the fitness and evolutionary success of variant viruses, leading to stronger theoretical foundations for selection of candidate vaccine viruses.

# Introduction

Antigenic evolution of human influenza A viruses is characterized by rapid drift, with structural changes in antigenic epitopes allowing the virus to escape existing immunity. Consequently, seasonal influenza continues to impose a major burden on human health causing 250,000 to 500,000 deaths annually (WHO 2009). Influenza vaccines, which remain the most effective means of disease prevention, currently comprise antigens from A(H1N1), A(H3N2) and B viruses predicted to elicit the most effective immune responses against circulating viruses in the forthcoming influenza season (WHO 2009; Klimov et al. 2012). The continually evolving antigenic phenotype of influenza A viruses presents an ongoing challenge for vaccine virus selection, as effectiveness is greatest when vaccine components are antigenically similar to circulating viruses. The HA glycoprotein is the key antigenic determinant of influenza viruses (Skehel & Wiley 2000) and consequently the most critical vaccine component. Neutralizing antibodies primarily bind to amino acids in protruding loops and helices that form defined antigenic sites (Wiley et al. 1981; Caton et al. 1982), and substitutions that alter epitope structure can inhibit antibody binding and help the virus escape existing immunity (Webster et al. 1982). When the selective advantage conferred is sufficient, novel antigenic variants will replace circulating viruses. Hence, phylogenies of influenza A HA sequences are characterized by the presence of a single predominant trunk lineage, and short side branches, representing the rapid turnover of the influenza virus population (Fitch et al. 1997; Nelson & Holmes 2007).

Antigenic changes in circulating influenza viruses are principally assessed by the HI assay (Hirst 1941; WHO 2011). Results of many HI assays can be summarized using cartographic approaches, which approximate antigenic dissimilarity by Euclidean distances between viruses and antisera on a map, with antigenic evolution in influenza represented as movement between clusters of viruses (Smith et al. 2004). The non-synonymous genetic mutation(s) causing transitions between antigenic clusters can be determined experimentally by reverse genetics (Koel et al. 2013), though this approach is often laborious, as multiple amino acid substitutions bridge each antigenic cluster transition, and individual substitutions need to be assessed. This approach recently demonstrated that transitions between antigenic clusters of H3N2 viruses are caused predominantly by single amino acid substitutions at positions near the receptor-binding site (Koel et al. 2013). However, major cluster transitions may not be the only antigenically important events and an exhaustive reverse genetics analysis of all observed substitutions is not feasible due to high levels of amino acid sequence diversity in HA (e.g. at 46% of amino acid positions, in this study).

An alternative approach is to integrate matching sequence, antigenic and 3-D structural data into models that allow us to attribute the observed antigenic differences in a dataset directly to their underlying causes. Reeve *et al.* developed such a model to identify surface-exposed regions of the capsid proteins of foot-and-mouth disease virus where substitutions were correlated with antigenic change, but were unable to show definitive causal connection with specific substitutions (Reeve et al. 2010). Various other computational approaches have similarly been used to identify antigenically important amino acid positions in influenza HA by comparison of predominant sequences of successive antigenic clusters and by comparing sequence and antigenic data (Smith et al. 2004; Lee & Chen 2004; Huang et al. 2012; Steinbrück & McHardy 2012; Sun et al. 2013).

In this paper, we: 1. extend the modeling approach of Reeve *et al.* (Reeve et al. 2010) to former seasonal influenza A(H1N1) viruses, focusing on these rather than A(H3N2) viruses, for which the role of neuraminidase-mediated agglutination of red blood cells (RBCs) has complicated the relationship between HI data and antigenic change (Lin et al. 2010), or the distinct A(H1N1)pdm09 viruses, which have largely remained antigenically similar since emerging in humans in 2009 (Barr et al. 2014); 2. attribute variation in HI titers to individual amino acid substitutions; 3. quantify their antigenic impact; 4. assess, by reverse genetics, the impact of a subset of the identified substitutions to validate the model; 5. show how inferences, based on the determinants of low-impact and high-impact antigenic changes, improve our understanding of the antigenic evolution of the virus; and 6. demonstrate that the characterization of these antigenic determinants allows us to accurately assess directly from HA gene sequence data the antigenicity of newly emerging viruses, measurement of which is critical to predicting the evolutionary success of newly emerging variants.

## Results

We compiled HI assay data from former seasonal H1N1 viruses isolated between 1997 and 2009, comprising 19,905 individual measurements of cross-reactivity between antisera, raised against 43 reference viruses, and 506 viruses. The HA1 sequences for all reference and test viruses in this HI dataset were used to generate a maximum clade credibility tree (Drummond et al. 2012). HA1 trees for the complete set of 506 viruses and for the 43 reference strains only are shown with corresponding HI titers represented as a heat-map in Figure 1.

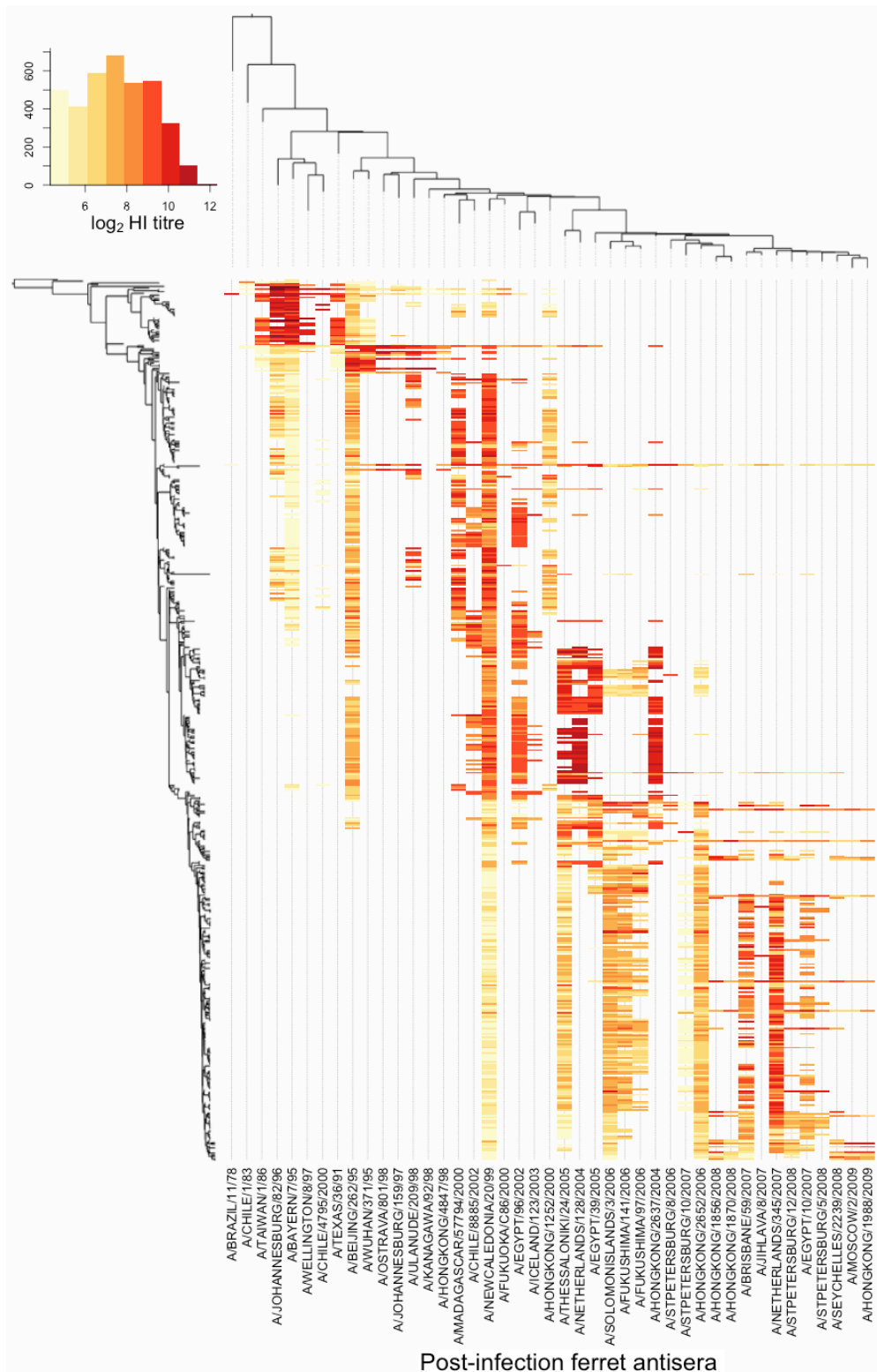

**Figure 1**. **Heat-map illustrating the relationship between molecular and antigenic evolution**. Cells are colored by mean log HI titer for each pairing of antiserum and test virus present in the full dataset. Test viruses and reference viruses used to generate post-infection ferret antisera are sorted phylogenetically on the HA gene along the vertical and horizontal axes respectively. Phylogenies are shown to the left for test viruses and above for reference viruses. The color key for HI titers is shown in the histogram at top left along with the number of assays yielding each titer. Figure S1 provides examples of the observed variability in HI titer for the most frequently used virus.

**The effect of amino acids at specific positions.** To identify antigenic relationships and their predictors, we used linear mixed effects models to account for variation in the HI titers, as described by Reeve *et al.* (Reeve et al. 2010). Initial model selection identified non-antigenic sources of variation in HI titer. We determined that a fixed effect, $a_v$, for each virus, *v*, should be included in the model (n=506, p<$10^{-20}$), to account for consistent differences in titers between viruses, reflecting changes in receptor-binding avidity amongst other factors. A further fixed effect, $s_r$, was required for each reference virus, *r* (n=43, p<$10^{-20}$), to account for consistent differences in titers between antisera raised against different reference viruses, potentially reflecting differences in immunogenicity. Date of test needed to be controlled for as a random effect, $\varepsilon_D$ (n=351, p<$10^{-20}$), accounting for variability in batches of RBCs and dilutions of RBCs, antisera and viruses. These factors compensate for non-antigenic effects impacting HI titers (Equation 1).

$$\log_2(H_{r,v}) = k_0 + k_1\alpha_1(r,v) + s_r + a_v + \varepsilon_D + \varepsilon_R$$

(1)

$H_{r,v}$ is the HI titer for test virus *v* and antiserum raised against reference virus *r*. $k_0$ is a baseline, and $\varepsilon_R$ is the residual measurement error not explained by the model.

Equation 1 includes a term, $k_1\alpha_1(r,v)$, to investigate the effect of amino acid substitutions at specific positions: $\alpha_1$ represents the presence (1) or absence (0) of substitution at a specific amino acid position between the reference virus, *r*, and test virus, *v*, and $k_1$ is the associated regression coefficient. Using this model (Equation 1) substitutions at over 50% of non-conserved, surface-exposed positions and over 25% of non-conserved, non-surface-exposed positions were significantly correlated with reduced HI titer (p<0.05) using a Holm-Bonferroni correction for multiple tests (Holm 1979). Furthermore, the number of synonymous mutations between viruses was significantly correlated with reduced titer (p<$10^{-15}$) because of a correlation between molecular and antigenic evolution. This demonstrates that a simple regression analysis will incorrectly identify some antigenically neutral changes as antigenically important – i.e. false positives – simply because they occur at a similar point in the evolutionary history of the virus to one or more antigenically important substitutions (*i.e.* in the same, or a nearby branch of the phylogeny).

**Incorporating phylogenetic structure.** The described tendency for identification of false positives required phylogenetic structure to be reflected in the model. Equation 7 of Reeve *et al.* (Reeve et al. 2010) was used to identify branches of the phylogeny that were correlated with lower HI titers when they separated reference virus and test virus:

$$\log_2(H_{r,v}) = k_0 + \sum_i m_i \delta_i(r,v) + s_r + a_v + \varepsilon_D + \varepsilon_R$$

(2)

Equation 2 incorporates branch terms $m_i \delta_i(r,v)$ instead of the term representing substitutions at a single amino acid position: $\delta_i = 1$ when reference virus ($r$) and test virus ($v$) are separated by branch $i$ of the phylogeny and $\delta_i = 0$ otherwise, with $m_i$ being the associated regression coefficient from the mixed effects model. The tree generated for the 506 viruses in our dataset contained 1010 branches and it was computationally unfeasible to search the $2^{1010}$ possible antigenically important sets of branches so a stochastic hill-climbing approach was used to identify 62 branches as correlating with drops in HI titer when they separated reference and test viruses, indicating that antigenic evolution occurred in these branches. Such antigenic events were found in much higher proportion in the trunk (38.3%) than in side (4.6%) branches ($\chi^2$ test, $p<10^{-13}$), supporting the standard model of influenza antigenic drift, whereby substitutions altering antigenicity without loss of fitness undergo preferential fixation, thus forming the trunk lineage from which future viruses descend (Fitch et al. 1997; Nelson & Holmes 2007). With these 62 branch terms included in the model (Equation 2), there was no longer a significant correlation between HI titer and the number of synonymous mutations between reference and test virus ($p>0.2$), nor with any of the non-surface-exposed positions ($p>0.05$). This shows that including branch terms accounted for the antigenic variability in the data and reduced the false discovery rate as expected.

**Substitutions affecting antigenicity in multiple positions of the phylogeny.** The model was extended by combining Equations 1 and 2 to include explicit terms for amino acid substitutions and branch terms, as in Equation 8 of Reeve *et al.* (Reeve et al. 2010), to identify antigenically important substitutions:

$$\log_2(H_{r,v}) = k_0 + k_1 \alpha_1(r,v) + \sum_i m_i \delta_i(r,v) + s_r + a_v + \varepsilon_D + \varepsilon_R$$

(3)

Equation 3 incorporates the term $k_1 \alpha_1(r,v)$ from Equation 1 representing substitution at specific amino acid positions. Each of the previously identified 62 branch terms ($\delta_i$) were included and associated regression coefficients ($m_i$) were re-estimated in a model containing the $k_1 \alpha_1(r,v)$ term. Because branch terms account for the antigenic changes inferred to occur in single specific branches of the phylogeny, any significant improvement to model fit by $\alpha_1$ is a result of the term representing amino acid substitution at a particular HA1 position being correlated with a change in the antigenicity of the virus represented in multiple branches of the phylogeny. Thus an improvement to model fit achieved by inclusion of α$_1$ indicates that there have been alternative, convergent- or back-substitutions at the same amino acid position associated with antigenic change in at least two branches of the phylogeny.

We investigated 113 non-conserved, surface exposed amino acid positions of the HA1 domain. At four of these (141, 153, 187 and 190), the inclusion of an α$_1$ term representing substitution (Equation 3) improved model-fit compared with the model containing only branch terms (Equation 2). Since the identified positions improve model fit in the presence of branch terms ($\delta_i$), we can infer that substitutions at these positions correlate with antigenic change in more than one position of the phylogeny. We describe substitutions at these positions as having *support across the phylogeny*. Each of these four amino acid positions (Figure 2A) has previously been allocated to one of the H1 antigenic sites (Caton et al. 1982). Position 187 is also a constituent of the primary sialic acid receptor-binding site and the analogous position 190 in H3-HA has been described as forming hydrogen bonds with the 9-hydroxyl group of sialic acid (Skehel & Wiley 2000). Non-surface exposed positions were examined separately; substitution at none of these positions improved model fit.

Different substitutions at the same position are expected to vary in antigenic impact according to the biochemical properties of the amino acids involved. To account for this, we measured the significance and average impact (in antigenic units, where a unit corresponds to a 2-fold dilution in the HI assay) of each substitution at HA1 positions 141, 153, 187 and 190 that was observed to have occurred between reference and test viruses in the dataset. Substitutions between seven pairs of amino acids at the four positions showed significant antigenic impact with *support across the phylogeny*. The mean antigenic impact ($k_1$ in Equation 3) of exchange between amino acids of each pair is shown in Table 1.

**Table 1. HA1 amino acid substitutions that correlate with antigenic change.**

| Substitution(s) (H1-HA numbering) | Antigenic site H1 (5) | Antigenic site H3 (4) | Antigenic impact * (antigenic units) | Support across the phylogeny |
|---|---|---|---|---|
| K141E | Ca | A | 2.37 (2.27-2.47) | Yes |
| E153G | Sa | B | 0.20 (0.07-0.33) | Yes |
| E153K | Sa | B | 0.66 (0.39-0.93) | Yes |
| G153K | Sa | B | 1.50 (0.51-2.49) | Yes |
| D187N | Sb | B | 0.33 (0.30-0.36) | Yes |
| D187V | Sb | B | 0.88 (0.51-2.49) | Yes |
| A190T | Sb | B | 0.24 (0.17-0.31) | Yes |
| S36N |  | C | 0.66 (0.22-1.11) | No |
| S72F | Cb | E | 0.81 (0.49-1.13) | No |
| E74G, E120G[†] | Cb,- | E,A | 0.43 (0.29-0.57) | No |
| R43L, F71I, A80V, ΔK130[†] | -,Cb, -,- | C,-, -,A | 3.53 (3.44-3.62) | No |
| S142N | Ca | A | 0.75 (0.58-0.92) | No |
| K163N | Sa |  | 0.67 (0.62-0.73) | No |
| S183P |  | B | 0.61 (0.33-0.89) | No |
| N184S | Sb | B | 0.51 (0.31-0.70) | No |
| W252R |  |  | 0.37 (0.32-0.43) | No |
| E274K |  |  | 1.31 (0.68-1.93) | No |
| R313K |  |  | 1.47 (0.84-2.10) | No |

* $k_1$ in Equation 3. Mean and 95% CI are shown. [†] Multiple substitutions in the same branch offer alternative explanations for the associated antigenic change.

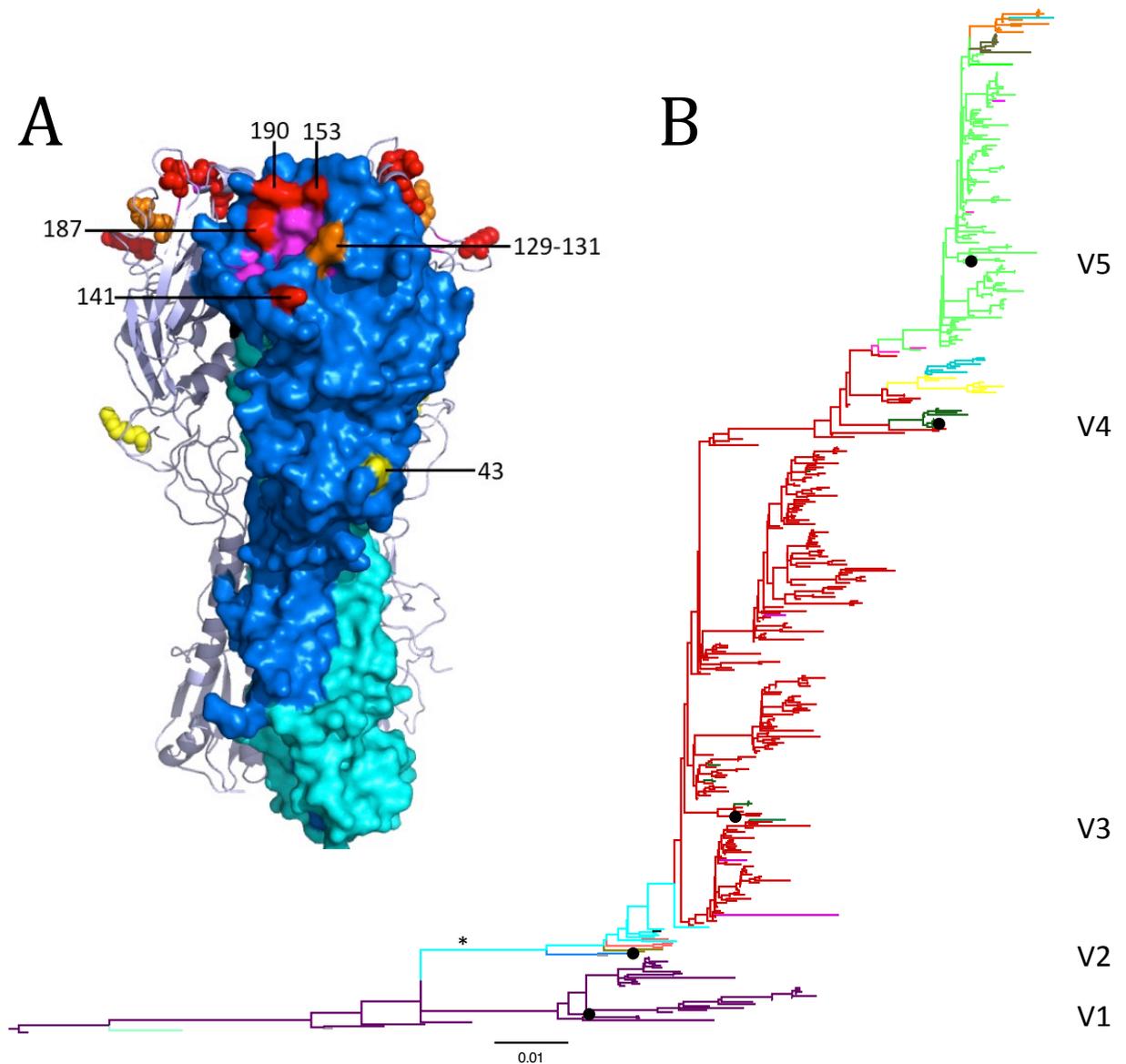

**Figure 2. HA positions implicated in antigenic evolution and locations of associated substitutions in HA1 phylogeny.** (**A**) Amino acid positions that affect antigenic phenotype modeled on the HA structure of A/Puerto Rico/8/34 (Protein Data Bank ID: 1RU7) (Gamblin et al. 2004). Surface representation of the front monomer is shown with HA2 in cyan and HA1 in blue with residues of the receptor-binding site colored pink. Positions with substitutions that can explain antigenic change in multiple locations across the phylogeny are shown in red. Residues adjacent to the position of the K130 deletion are colored orange with the locations of the co-occurring R43L, F71I and A80V substitutions are colored yellow. Residues are labeled on the front HA1 monomer and shown as spheres on the remaining backbones. (**B**) HA1 phylogeny showing positions of significant antigenic substitutions. Color changes mark the locations of substitutions associated with changes in antigenic phenotype of at least 0.5 antigenic units. The position of the branch associated with the greatest drop in cross-reactivity is marked (*). Black circles indicate the positions of viruses included in the influenza vaccine over the period of HI data collection and are labeled: A/Bayern/7/95 (V1), A/Beijing/262/95 (V2), A/New Caledonia/20/99 (V3), A/Solomon Islands/3/2006 (V4) and A/Brisbane/59/2007 (V5). Branch length indicates the estimated number of nucleotide substitutions per site.

**Substitutions affecting antigenicity at single positions in the phylogeny.**
Next, we added terms $k_j$ and $α_j$ to represent each of the seven inferred antigenic substitutions at the four positions with support across the phylogeny (Table 1) to produce Equation 4. We then investigated the causes of antigenic change in branches that still had large estimated antigenic impacts.

$$\log_2(H_{r,v}) = k_0 + k'α'(r,v) + \sum_j k_jα_j(r,v) + \sum_i m_iδ_i(r,v) + s_r + a_v + ε_D + ε_R$$

(4)

Terms for these seven substitutions absorb variation in HI previously explained by branch terms that correspond to the positions in the phylogeny where those substitutions were estimated to have occurred. However, this model still included 18 branch terms representing internal branches of the phylogeny whose estimated impact, $m_i$, on the HI assay remained detectable (at least 0.25 antigenic units) in the model containing terms for each of the seven substitutions. Each of these 18 branch terms were excluded in turn, the model re-built with the residual branches, and each remaining amino acid position (as $k'$ and $α'$) was retested to determine which substitution(s) could explain the variation in HI titer associated with the excluded branch term. A substitution identified at this stage (when a branch term had been excluded) was inferred to have caused the associated antigenic change at that position in the phylogeny if it was the only substitution identified.

In nine cases, a single substitution was identified as explaining variation in HI titer upon exclusion of one branch. These substitutions were at positions 36, 72, 142, 163, 183, 184, 252, 274 and 313 (Table 1). Unique identification was not possible in two further cases, as multiple substitutions occurring in the same branch could not be discriminated. The branch associated with the greatest drop in HI titer (-3.61 units) across the phylogeny (starred in Figure 2B) has a deletion of lysine at position 130 (ΔK130) and substitutions R43L, F71I and A80V. The antigenic significance of ΔK130 has been described (McDonald et al. 2007); however each of these co-occurring substitutions have been identified as antigenic determinants by another *in silico* technique, which did not identify them as false positives (Huang et al. 2012). Each of these four changes is assigned equal weight in our model but we identify explicitly that they offer alternative explanations for the same antigenic change and are not independent antigenic determinants. To infer their individual effects experimental investigation was required. One further instance of alternative substitutions at different positions explaining an antigenic change equally well involved positions 74 and 120. The nine single substitutions and these two instances where alternative substitutions explained antigenic change equally well gave a total of eleven cases where antigenic change in a single position of the phylogeny could be attributed to an amino acid substitution, or multiple substitutions (Table 1).

Although the substitutions identified when branch terms at these eleven positions of the phylogeny were excluded correlated with antigenic change at only a single position in the phylogeny, it is notable that, among them, positions 72, 74, 142, 163 and 184 map to previously described H1 antigenic sites while positions analogous to 36, 120 and 183 are constituents of H3 antigenic sites (Wiley et al. 1981). Locations within the phylogeny where any of the identified substitutions in Table 1 altered the antigenic phenotype of the virus by at least 0.5 antigenic units and the degree of correspondence with changes to the H1 vaccine component are shown in Figure 2B. Each of the five vaccine components in this phylogeny are separated by at least one color change indicating that potential genetic drivers for all of the most important antigenic changes in the period studied have been identified.

**Production of mutant viruses by reverse genetics.** To validate the identification of substitutions affecting antigenicity and assess the accuracy of estimated antigenic effects mutant viruses containing a subset of the amino acid substitutions identified in Table 1 were generated by reverse genetics. The HA gene of an exclusively cell culture-propagated virus, A/Netherlands/1/93 (Neth93), was used. We introduced the K130 deletion (ΔK130) and the R43L substitution into the Neth93 HA independently to test whether both of these changes cause antigenic change. Given the large antigenic impact of ΔK130 (McDonald et al. 2007), its introduction generated an additional, antigenically distinct HA background (Neth93 Δ130) in which to further test the effects of other substitutions (Table 1): the HA genes of both Neth93 and Neth93 Δ130 were used to produce viruses carrying individual substitutions of K141E, E153K and D187N.

Mutant recombinant viruses were characterized by HI using a panel of post-infection ferret antisera raised against seven reference viruses with HA1 amino acid identity at positions 43, 130, 141, 153 and 187 as shown in Table S1. To assess the antigenic impact of each amino acid substitution introduced by mutagenesis, antisera of two types were chosen: 1) antisera raised against parent-like viruses that had amino acid identity in common with the parent virus (i.e. R43 for the substitution R43L); 2) antisera raised against mutant-like viruses that had amino acid identity in common with the recombinant virus (i.e. L43 for the substitution R43L). For each amino acid substitution introduced into the recombinant viruses, the assignment of antisera to these two categories, based on reference virus amino acid sequence, is shown in red or blue cell color respectively in Table S1.

**Estimating the antigenic and non-antigenic effects of introduced substitutions.** In addition to antigenic change, HI titers can be affected by variation in other properties of the test virus, notably receptor-binding avidity, and individual amino acid substitutions may cause fluctuation in HI titer as a result of variation in these other properties (Daniels et al. 1984; Hensley et al. 2009). By using antisera raised against parent-like and mutant-like viruses, changes in log$_2$ HI titer between parent and mutant recombinant viruses resulting from antigenic ($\Delta H_A$) and non-antigenic ($\Delta H_N$) effects could be distinguished using Equation 5.

$$\Delta H_A = \frac{(\Delta H_2 - \Delta H_1)}{2} \qquad \Delta H_N = \frac{(\Delta H_1 + \Delta H_2)}{2}$$

(5)

If the amino acid substitution introduced into the parent virus was antigenically important it was expected to cause a decrease in HI titer for the mutant virus against antisera induced by parent-like virus ($\Delta H_1$, Equation 5) and a corresponding increase against antisera raised against mutant-like virus ($\Delta H_2$, Equation 5). Conversely, a change in virus receptor-binding avidity is expected to cause a consistent decrease (or increase) in titer with these two groups of antisera ($\Delta H_1$ and $\Delta H_2$). Therefore, for each substitution the associated change in log$_2$ HI titer, relative to Neth93 or Neth93 Δ130, were partitioned into antigenic ($\Delta H_A$) and non-antigenic ($\Delta H_N$) components.

The predicted antigenic effect of each substitution (from Table 1) is shown alongside mean observed changes in log$_2$ HI titer partitioned into antigenic ($\Delta H_A$) and non-antigenic effects ($\Delta H_N$) in Table 2. The range of antigenic effects of K141E, ΔK130, E153K and D187N amino acid substitutions, measured against the panel of antisera, were consistent with predictions from the modeling. The range in antigenic impact ($\Delta H_A$) measured using individual antisera is shown in Figure 3 and the mean observed titers averaged across four repeats are shown in Table S2 and as a heat-map in Figure S2. Across all substitutions we observed a mean error in our predictions of 0.14 antigenic units.

**Table 2. Comparison of predicted and observed antigenic impacts of HA1 amino acid substitutions assessed by HI.**

| Substitution | Predicted antigenic effect* | Mutagenesis Background | Observed effect[†] $\Delta H_A$ | $\Delta H_N$ |
|---|---|---|---|---|
| K141E[‡] | 2.37 | Neth93 Δ130 | 2.60 | +0.27 |
| E153K | 0.66 | Neth93 | 0.67 | -0.42 |
| | | Neth93 Δ130 | 0.65 | **-2.15** |
| | | Averaged | 0.66 | **-1.28** |
| D187N | 0.33 | Neth93 | 0.41 | -0.41 |
| | | Neth93 Δ130 | -0.08 | -0.54 |
| | | Averaged | 0.16 | -0.47 |
| ΔK130 | 3.53[§] | Neth93 | 4.10 | -0.78 |
| R43L | 3.53[§] | Neth93 | 0.01 | -0.01 |

* Predicted mean antigenic impact (from Table 1) measured in antigenic units. [†] Mean observed changes in log$_2$ HI titer (in antigenic units) partitioned into antigenic ($\Delta H_A$) and non-antigenic ($\Delta H_N$) effects. [‡] HA plasmid was generated for the mutant Neth93 K141E but multiple attempts to rescue virus were unsuccessful. [§] ΔK130 and R43L occur in the same branch of the phylogeny offering alternative explanations for the associated antigenic change.

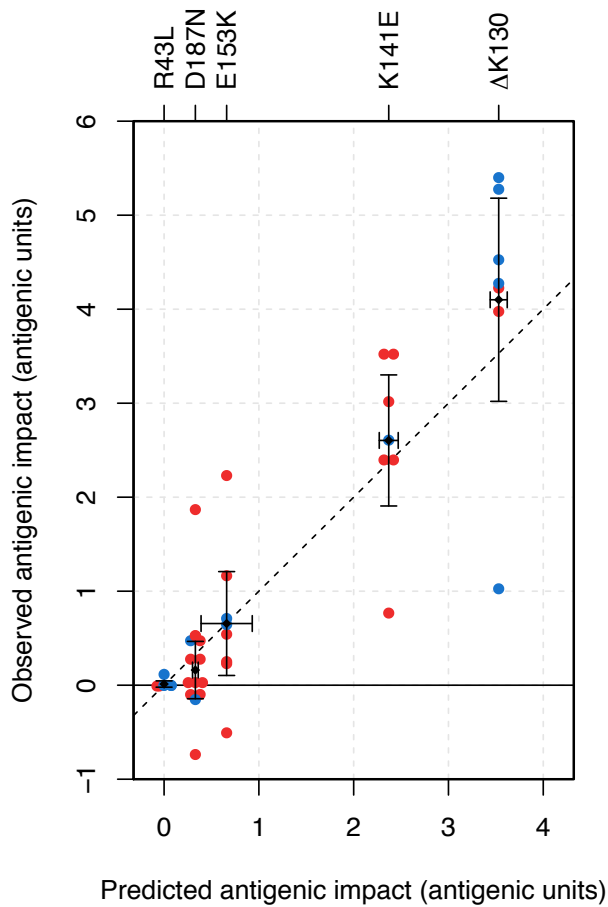

**Figure 3. Observed and predicted antigenic impact of amino acid substitutions.** The mean antigenic impact of each substitution predicted from modeling (Table 1) plotted against the mean observed impact averaged across antisera in the panel (Table S1). 95% confidence intervals are shown for both. Each point shows the observed mean antigenic impact ($\Delta H_A$, change in HI titer for a recombinant virus relative to its parent virus) of a particular amino acid substitution (labeled at top) with each antiserum in the panel. Red points indicate that the reference virus lacked the amino acid substitution, so the predicted impact of mutation is a reduction in titer; blue points indicate that the reference virus shared the substitution, so the predicted impact of mutation is an increase in titer. The number of points for each substitution is dependent on whether it was inserted into one or both (Neth93 and Neth93 Δ130) parental viruses and on the number of antisera used. A negative observed antigenic impact indicates a change in HI titer in the opposite direction to that predicted. Mean titers used to calculate antigenic and non-antigenic effects of substitutions are shown in Table S2 and as a heat-map in Figure S2.

The predicted and observed antigenic impacts, based on HI results, shown in Table 2 and Figure 3 indicate that our model captures the mean impacts of the HA1 amino acid substitutions identified. However, we also observed non-antigenic effects ($\Delta H_N$) of substitutions that resulted in higher or lower HI titers irrespective of antigenic similarity between test virus and the reference virus against which a particular antiserum was raised. Such effects exceeding 0.78 antigenic units, shown in bold in Table 2, cannot be explained solely by differences in virus concentration resulting from the limited accuracy of the hemagglutination assay, used to standardize hemagglutinating units prior to HI. We observed a relatively small antigenic impact of the E153K substitution, but a large non-antigenic effect, on HI titer. This result is supported by previous work showing E153K to have a relatively small impact on monoclonal antibody binding while causing a large increase in receptor-binding avidity that together contributed to large reductions in HI titer indicating escape from inhibition by polyclonal antiserum (Hensley et al. 2009).

**Antigenic cartography.** Among the substitutions we identified, the mean antigenic effect of two, ΔK130 and K141E, was estimated to be greater than two antigenic units. Each of these substitutions has been previously identified as antigenically important (ΔK130 (McDonald et al. 2007), K141E (Koel et al. 2013)), and K141E has been associated with a transition between clusters of antigenically distinct viruses in the antigenic evolution of H1N1 (Koel et al. 2013). To visualize antigenic evolution, viruses and antisera were positioned in a two dimensional antigenic map using a Bayesian multidimensional scaling model that estimates reference virus immunogenicity and test virus receptor-binding avidity (Bedford et al. 2014). Examination of the resulting antigenic map showed that the substitutions ΔK130 and K141E could explain the two transitions between antigenic clusters that occurred during the period of H1 evolution studied (Figure 4), and were the only antigenic substitutions identified that separated these clusters.

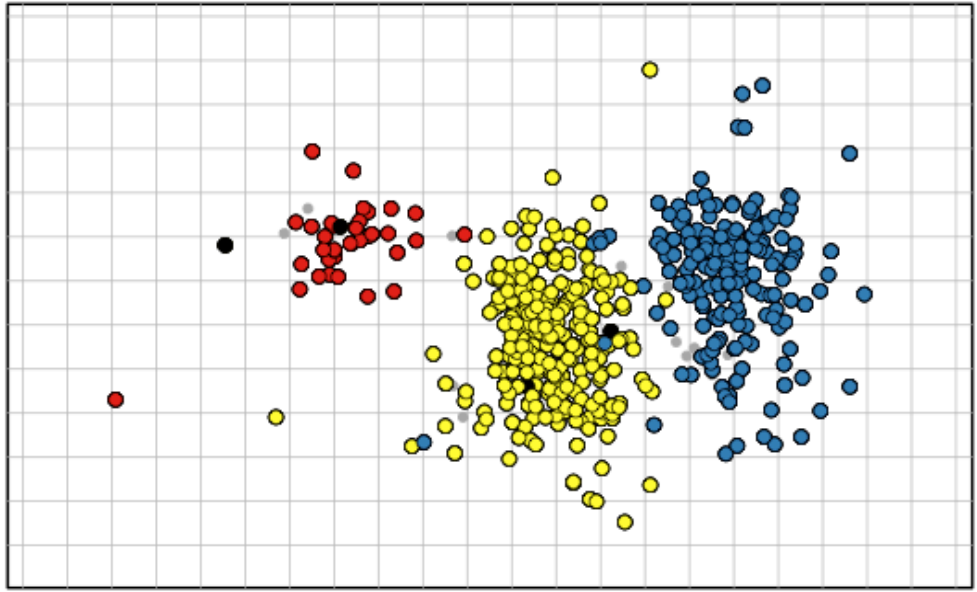

**Figure 4. Position of substitutions ΔK130 and K141E on an antigenic map**. Map locations are shown for a representative example from a Bayesian multidimensional scaling model that estimates virus location, antiserum location, reference virus immunogenicity and test virus receptor-binding avidity. Gridlines represent single antigenic units, twofold dilutions in the HI assay. Viruses are shown as colored circles and antisera are shown as grey points. Viruses are colored in relation to the substitutions ΔK130 and K141E: 130K 141K (red, n=36), Δ 130 141K (yellow, n=273), Δ130 141E (blue, n=193). Viruses with other amino acid combinations are colored black (n=4).

**Sequence-based prediction.** Finally**,** to assess whether the 18 inferred antigenic determinants shown in Table 1 were predictive of antigenic phenotype, as assessed by HI titer, our results were cross-validated 100 times using model parameters derived from 90% of the data, randomly selected each time. These 18 included two cases with multiple, co-occurring HA substitutions (R43L, F71I, A80V with ΔK130, and E74G with E120G) and 16 single substitutions. ΔK130 was selected rather than the co-occurring substitutions R43L, F71I or A80V given the results of genetics experiments described in Table 2 and Figure 3, however a single ambiguous term was used to represent either E74G or E120G as these have not been discriminated between.

$$\log_2(H_{r,v}) = k_0 + \sum_{\omega \in \Omega} m_\omega \times \alpha_\omega(r,v) + s_r + a_v$$

(6)

Equation 6 describes the predictive model, based on Equation 1, that estimates the antigenic dissimilarity of two viruses based on which substitutions separate them. $\alpha_\omega$ is 1 when reference virus ($r$) and test virus ($v$) are separated by a specific substitution (or its reverse), and 0 otherwise. The substitutions included in the model are Ω = {S36N, S72F, E74G or E120G, ΔK130, K141E, S142N, E153G, E153K, G153K, K163N, S183P, N184S, D187N, D187V, A190T, W252R, E274K, R313K} – all of the substitutions identified above. Each substitution in Ω also has an associated antigenic impact, $m_\omega$, previously identified in Table 1, but here estimated repeatedly in the model from (90%) training data to predict the (10%) test data in a cross-validation procedure. We compared the prediction error of this model with and without the parameters $a_v$ and $s_r$ to investigate the importance of including non-antigenic effects in the model, and also with the subset of Ω containing only substitutions that defined the clusters found on the antigenic map (Ω′ = {ΔK130, K141E}) to investigate the importance of low-impact substitutions.

A simple null model that contained no substitution terms and that used only the average titer for antiserum raised against each reference virus ($s_r$) to predict antigenic phenotype, produced a mean absolute error of 1.32 units (Figure 5A). By including only identified cluster-defining substitutions during the period studied, ΔK130 and K141E, prediction improved, with a mean absolute error of 1.02 antigenic units (Figure 5B). Adding in all 18 substitution(s) (Table 1) reduced this to an error of 0.69 units (Figure 5C). Inclusion of lower-impact, non-cluster defining substitutions therefore allowed more accurate prediction of antigenic phenotype. Prediction accuracy was further improved by compensating for non-antigenic differences between viruses (i.e. variation in their receptor-binding avidity). Allowing the average titer for each virus to absorb this variation and using $a_v$ terms for prediction resulted in an error of only 0.54 antigenic units (Figure 5D). When a virus that is not present in the training data appears in the test data the reactivity parameter ($a_v$) associated with that virus is set to the mean of the training virus reactivities.

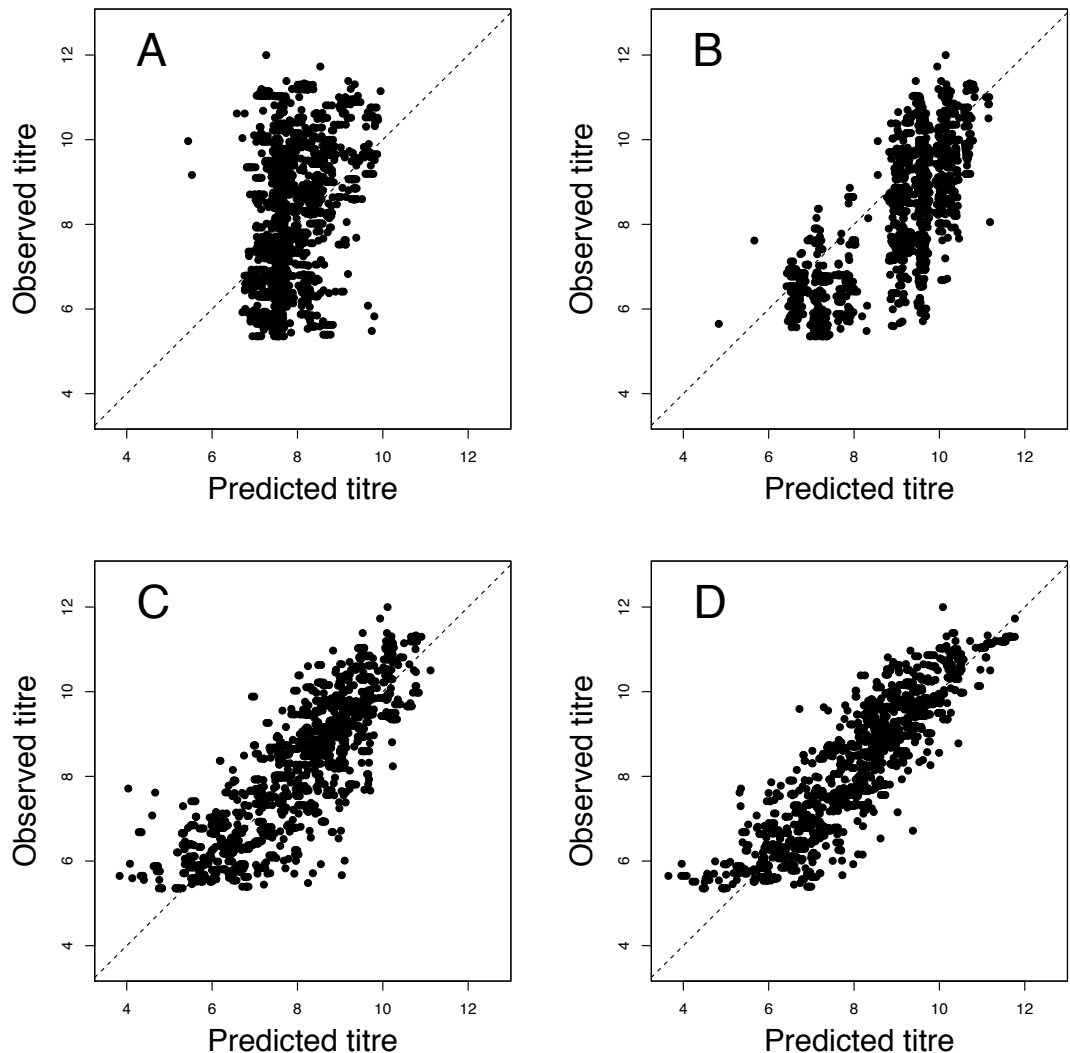

**Figure 5. Sequence-based prediction of antigenic phenotype.** Observed and predicted HI titers plotted on $\log_2$ scale (antigenic units) using representative models trained with 90% of the data. Predictive models contained terms for A) Average titers for each reference virus, B) Antigenic cluster-defining substitutions ΔK130 and K141E, C) All 18 antigenic substitution(s) shown in Table 1, D) All 18 antigenic substitution(s) shown in Table 1 with additional terms that estimate differences in test virus receptor-binding avidity (non-antigenic variation in titer associated with each virus). Each model was fitted to the same training dataset comprising 90% of all observations and predictions for the remaining data are shown. Incremental improvements in mean absolute prediction error are shown alongside SEM and 95% upper limits in Table S3.

To further investigate the predictive power of the approach, the antigenic phenotype of viruses isolated in each year of our dataset (from 1998 to 2009) was predicted using model parameters derived using only data collected prior to that year. We modeled the HI titers of each test virus isolated in a year using antisera raised against viruses isolated in all previous years. Mean absolute prediction error averaged across the twelve years was 1.81 units (SEM=0.0112, SD=0.92) when only the substitutions ΔK130 and K141E were included. This reduced to 0.90 units (SEM=0.0085, SD=0.70) when all 18 substitutions were included. The mean absolute prediction error of each model in each year is shown in Figure 6; in all years the accuracy is improved by the inclusion of the lower-impact antigenic determinants in addition to the two substitutions that define antigenic clusters on the map.

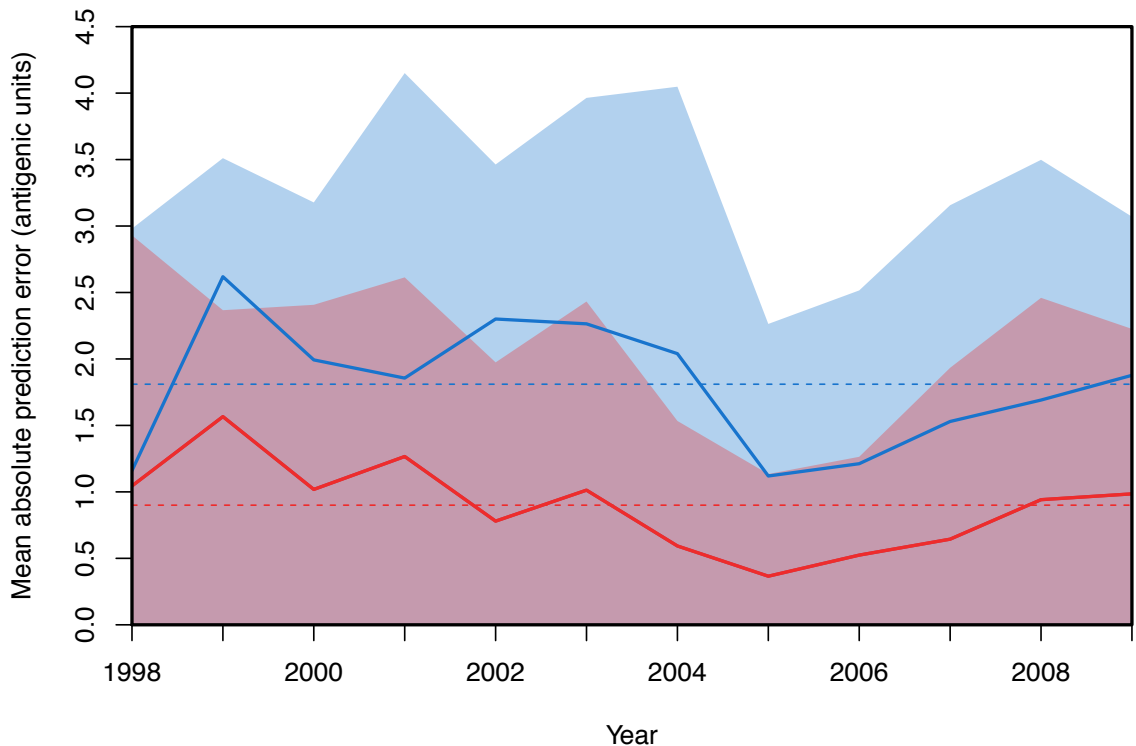

**Figure 6. Prediction error through time for models used to predict HI titers of viruses isolated in the following year.** The mean, absolute difference between observed titers for viruses isolated in a given year and titers predicted using models trained to HI data collected in previous years is shown. Predictive models included terms for cluster-defining substitutions ΔK130 and K141E only (solid blue line) or for all 18 substitutions in Table 1 (solid red line). For each model, shaded areas show the lower 95% confidence interval on the absolute prediction error and the mean, absolute prediction errors averaged across the twelve years is shown as a dashed line.

## Discussion

Using a modeling approach that integrated HA sequence data and HI antigenic data for over 500 viruses, we have identified substitutions responsible for the antigenic evolution of former seasonal influenza A(H1N1) viruses over a period of more than 10 years. We identified substitutions at 15 amino acid positions in HA1: two that had high-impact on antigenicity (which individually can lead to a need to change vaccine virus), and 13 of lower impact, including some too low to be directly observable in routine HI assays. Substitutions at four of the fifteen amino acid positions occurred multiple times in the evolutionary history of the virus, consistently explaining observed antigenic changes. Antigenic cartography of the H1N1 viruses identified only three antigenic clusters, transitions between which can be explained by the two high-impact substitutions.

As here with H1N1, in antigenic maps of H3N2 viruses antigenic distances between viruses belonging to the same antigenic cluster often exceeded distances between viruses in adjacent clusters, demonstrating the need to assess non-cluster defining substitutions (Smith et al. 2004). Further, the selection of approximately twice as many H3N2 vaccine viruses as antigenic clusters identified by Smith *et al.* during the period 1968 to 2003 is supportive of the importance of substitutions not readily identified using antigenic maps (Koel et al. 2013). Using cross-validation we show that inclusion of substitutions causing low- to high-impact antigenic changes significantly improved prediction for the viruses of the H1N1 subtype. Furthermore, when predicting the antigenic phenotype of viruses in the following year, including these substitutions doubled accuracy; in contrast, using only cluster-defining substitutions generated worse predictions in each year and in several years mean absolute prediction error for these models exceeded two antigenic units. An antigenic distance of two units may necessitate a change of vaccine virus recommendation, so this improvement has significant implications for the usefulness of these predictive models. The improved accuracy of predictions made using the full model shows that substitutions causing low-impact antigenic changes are common in the evolution of influenza A viruses and crucial to the tracking of antigenic evolution. Identification of substitutions responsible for such smaller incremental changes in antigenicity also raises the prospect of fine-tuning vaccine viruses by mutating existing candidate vaccine viruses or their derivatives.

The model also partitions the results of the HI assay into antigenic and non-antigenic effects, and we have done the same when characterizing the mutant viruses generated by reverse genetics. At position 153 we detect a non-antigenic effect of substitution contributing to apparent antigenic effects in HI titers, consistent with previous studies of former seasonal H1N1 (Hensley et al. 2009). Substitutions in the 150-loop (153-157) of HA1 have been shown to occur during culture of A(H1N1)pdm09 viruses (Klimov et al. 2012) and G155E substitution has been shown to affect receptor-binding specificity or avidity (Liu et al. 2010): such receptor-binding alterations probably contribute to apparent antigenic effects attributed to substitutions introduced into this region of the A(H1N1)pdm09 HA1 by reverse genetics (Koel et al. 2015). Understanding the genetic variation underlying changes in the receptor-binding avidity of influenza viruses that contribute to apparent antigenic effects as measured by HI assay is clearly a very important area for further investigation. We have also not explicitly modeled how the antigenic impact of substitution at one position depends on substitutions present at other HA, or neuraminidase, positions. It has been proposed that epistasis is prevalent in the evolution of influenza surface glycoproteins, however few examples have been confirmed using phenotypic data (Kryazhimskiy et al. 2011). In the assessment of viruses derived by reverse genetics performed here, the effect of substitutions varied with the antiserum used in HI assays. Understanding how variation in the antigenic impact of a particular substitution can be attributed to existing genetic and antigenic differences between viruses is another target for future research that should further improve our understanding of how specific substitutions affect antigenicity.

Łuksza and Lässig recently demonstrated that they can predict the evolutionary success of influenza clades using population genetics models that incorporate information on the frequencies of genotypes in the previous season and the number of substitutions in known antigenic and non-antigenic regions (Łuksza & Lässig 2014). However, the effectiveness of such an approach depends on the ability to quantify how individual mutations affect fitness (Koelle & Rasmussen 2014). Antigenicity is a key component of the fitness of human influenza viruses, and we show that we can both quantify the impact of specific amino acid changes and use this knowledge to predict antigenic phenotype directly from sequence data. Our ability to quantify heterogeneities in the antigenic impact of substitutions improves our understanding of the genetic basis of antigenic evolution in influenza viruses. This allows sequence-based predictions of antigenicity to be made for genetic variants before HI testing, thereby increasing the value of sequence data from emerging genetic variants to assist targeting of antigenic analyses and hasten the identification of emergent antigenic variants. We anticipate that the ability to determine roles of both high- and low-impact amino acid substitutions in antigenic drift will complement existing methods and improve genotype-based predictions of virus fitness and consequent evolutionary success.

## Materials and Methods

**Data.** Viruses were originally isolated from clinical specimens either by WHO National Influenza Centres (NICs) or by the WHO Collaborating Centre. The antigenic dataset encompassed 506 former seasonal A(H1N1) viruses for which HA gene sequence data were available, inclusive of 43 reference viruses against which post-infection ferret antisera were raised, with 19,905 HI titers measured between 3,734 unique combinations of virus and antiserum, made on 351 dates from 1997 to 2009.

**Recombinant viruses.** Viruses were generated using a protocol based on Hoffman *et al.* 2000 (Hoffmann et al. 2000). HA and neuraminidase cDNAs of A/Netherlands/1/93 (Neth93), which had been exclusively propagated in cell culture, were amplified using a standard RT-PCR protocol. These cDNAs were cloned into the pHW2000 vector. Mutations were introduced into the HA plasmid using the QuikChange lightning site-directed mutagenesis kit (Agilent Technologies, Santa Clara, California). Co-cultured 293T and MDCK cells were co-transfected with plasmids containing HA and neuraminidase derived from Neth93 with the remaining six genes from A/Puerto Rico/8/34. After 2-3 days, recombinant viruses in the supernatant of transfected cells were recovered and propagated in MDCK cells as described in Lin *et al.* (Lin et al. 2012). Virus HA sequences were verified after passage.

**HI assays and analysis**. HI assays were performed on recombinant viruses by standard methods (WHO 2011). Post-infection ferret antisera raised against the following reference viruses were used: A/Bayern/7/95, A/Johannesburg/82/96, A/Johannesburg/159/97, A/Ulan-Ude/209/98, A/Hong Kong/4847/98, A/New Caledonia/20/99 and A/Hong Kong/1252/2000. Amino acid identities at HA1 positions 43, 130, 141, 153 and 187 of these seven reference viruses are shown in Table S1.

Average changes in $\log_2$ HI titer between parent and mutant recombinant viruses were quantified and partitioned into antigenic ($\Delta H_A$) and non-antigenic ($\Delta H_N$) effects using Equation 5. Antigenic effects were compared with predictions from modeling. Mean error in predictions across all substitutions was calculated as the average difference between the predicted mean and each measured antigenic change in HI using a specific virus dilution measured against a particular antiserum (excluding measurements restricted by the lower threshold of the HI assay).

Small non-antigenic changes in HI titer ($\Delta H_N$) between two viruses could be explained by the routine standardization of both viruses using the hemagglutination assay prior to HI. Limitations in the accuracy of the hemagglutination assay controlling for virus concentration ($\pm$ 0.5 hemagglutinating units) for both parent and mutant viruses mean that effects on HI titer below 0.78 antigenic units (95% CI) could be a result of test error. Corresponding antigenic changes, however, look at differences between antisera for a single sample of the same diluted virus, controlling for this effect.

**Phylogenetic analysis.** HA1 nucleotide sequences of the 506 viruses were aligned using MUSCLE (Edgar 2004). Phylogeny construction and analysis was carried out using BEAST v1.7.4 (Drummond et al. 2012) which uses Markov chain Monte Carlo (MCMC) to explore parameter space and evaluate phylogenetic models and Tracer v1.5 (Rambaut & Drummond 2009). Phylogenies were estimated using a variety of nucleotide substitution and molecular clock models. A relaxed, uncorrelated clock and a GTR+I+$\Gamma_4$ nucleotide substitution model were determined to be most suitable through comparison of Bayes factors (Suchard et al. 2001). Bayes factor analysis also determined that a separate partition should be created for the third codon position to allow rates of nucleotide substitution at this position to vary relative to the first and second codon positions. As a prior, we assumed an underlying coalescent process with a constant population size on the tree. Time of isolation was used to calibrate the molecular clock allowing rates of evolution along branches to be estimated. The maximum clade credibility tree was identified from a posterior sample of 10,000 trees. Substitution at position 187, associated with adaptation to propagation in eggs (Raymond et al. 1986; Robertson et al. 1987; Gambaryan et al. 1999), was assumed to be an artifact with potential to distort phylogenetic inference, so nucleotides coding for position 187 were excluded from phylogenetic analysis. Ancestral amino acid state at each node in the phylogeny for each position identified by modeling was estimated using the FLU amino acid substitution model (Dang et al. 2010) and unlinked strict molecular clocks for each amino acid position.

**Antigenic cartography.** Virus locations in antigenic space were estimated using the Bayesian multidimensional scaling technique of Bedford *et al.* (Bedford et al. 2014), which extends Smith *et al.* (Smith et al. 2004) by incorporating a phylogenetic diffusion process and estimates of antiserum and virus reactivity to account for variation in the immunogenicity of different reference viruses and in the receptor-binding avidity of viruses.

**Mixed effects modeling and model selection.** Co-variance between HI titer and the size of residuals from models using linear HI titers necessitated the use of logarithmically transformed HI titer as the response variable (to ensure homoscedasticity). Base 2 was chosen (without loss of generality) for the logarithm to follow Smith *et al.* and work throughout was in terms of $\log_2$ (or *antigenic units*) where 1 corresponds to a two-fold dilution of antiserum in the HI assay (Smith et al. 2004). The likelihood ratio test was used to test models containing combinations of the following explanatory variables: the reference virus against which the antiserum was raised, the test virus, and the date on which the assay was performed.

Following Reeve *et al.* (Reeve et al. 2010), each branch of the HA1 phylogeny was tested in the model as a fixed effect term. Each branch term was included as a discrete indicator variable: 1 when reference virus and test virus were separated by the branch in the phylogenetic tree and 0 otherwise. Random restart hill-climbing was used to determine the best model (Russell & Norvig 1995). To a random consistent starting model, branch terms were added and removed at random to maximize model fit, assessed by AIC (Akaike 1974). This was repeated while randomizing their order to identify the best model to avoid sensitivity to the order in which the parameters were presented. This approach was conservative since it was used to determine the branches used to control for phylogenetic correlations in the data, and adding in extra unnecessary terms simply reduced the power of the analysis. We show elsewhere that sparse Bayesian variable selection methods are more powerful for such problems, but are not computationally feasible for use on such large datasets (Davies et al.

2014). Conceptually similar machine learning techniques have been used on related influenza datasets (Sun et al. 2013), but these did not control for phylogenetic correlation, which we consider to be critical to avoid false positives, and this remains a computationally expensive step.

MacPyMOL Molecular Graphics System v1.3 was used to visualize and identify surface-exposed positions on the HA 3D-structure of A/Puerto Rico/8/34 (resolved to 2.3Å, Protein Data Bank ID: 1RU7) (Gamblin et al. 2004). To account for potential structural changes during the period of evolution since the isolation of A/Puerto Rico/8/34, the HA 3D-structure of A/Solomon Islands/3/2006 (resolved to 3.19Å, Protein Data Bank ID: 3SM5) (Whittle et al. 2011) was also used. Amino acid dissimilarity between reference virus and test virus at each position exposed on the surface of either structure, not conserved within the dataset, was tested as a predictor of reduced HI titer ($p<0.05$) using a Holm-Bonferroni correction to account for multiple tests (Holm 1979). Although position 187 was excluded from phylogenetic inference, amino acid dissimilarity at this position was tested as a predictor of antigenic difference. At each HA position identified at this stage, the mean antigenic impact of specific amino acid substitutions was determined by examining the associated parameter ($k_1$, Equation 3) using data subsets with only two amino acid variants at that position.

Amino acid positions at which substitution correlated with antigenic change at only a single position in the phylogeny were identified next. Branches retained by the random restart hill-climbing approach, but not correlated with any substitutions that explained antigenic change in multiple branches, were identified by including terms for those substitutions in the model, and examining the resultant regression parameter ($m_i$, Equation 4) associated with each branch. Branch terms whose effects remained detectable ($m_i > 0.25$) were removed sequentially. Amino acid dissimilarity between reference virus and test virus at each non-conserved surface-exposed position was re-tested for inclusion in the model in the absence of each branch term. Statistical analyses were performed using R software (R Core Team 2012) and the package lme4 (Bates et al. 2012).

**Sequence-based prediction.** To test the predictive power of the identified substitutions, repeated randomized cross-validation was used: 90% of the data was repeatedly selected at random (100 times) to act as the training dataset to which models were fitted allowing predictions of titer to be made for the test dataset composed of the remaining 10% of the data. Branch terms included in the antigenic site analysis, to prevent identifications of false positives, were not included in these predictive models since it is only the substitutions themselves that are causally connected to antigenic change.

HI titers are affected by a number of experimental variables and thus observed titers are affected by substantial experimental variability as illustrated in Figure

S1. Therefore, we compared predictions with the estimated underlying average titers between antiserum ($r$) and test virus ($v$), fitted using the linear mixed-effects model that best fitted the data (Equation 7). In addition to fixed effects for antiserum ($s_r$), test virus reactivity ($a_v$) and the interaction between them ($\gamma_{rv}$, which estimates their antigenic relationship directly), $\varepsilon_D$ and $\varepsilon_R$ encapsulate a random effect for date and residual measurement error, respectively.

$$\log_2(H_{r,v}) = k_0 + s_r + a_v + \gamma_{rv} + \varepsilon_D + \varepsilon_R$$

(7)

Additional to using 10% of the data as a test dataset, we examined the ability of the same models to predict antigenic relationships between existing antisera and 'future' viruses. Test datasets containing all observations between viruses isolated in a given year and antisera raised against reference viruses isolated in all previous years were constructed. Data from all previous years were used as training datasets. Models were implemented using JAGS v3.3.0 (Plummer 2012) through R using the package runjags (Denwood 2013).

## Acknowledgements

We acknowledge the network of WHO National Influenza Centres and WHO Collaborating Centres that comprise the WHO Global Influenza Surveillance and Response System, who provided the influenza viruses used in this study. We specifically thank Dr. J. de Jong, Erasmus Medical Centre, Rotterdam, for providing A/Netherlands/1/93. This research was supported by the Medical Research Council under programme number U117512723 (WTH, VG, DJB, RSD, AJH, JWM) and grant MR/J50032X/1 (WTH), the Wellcome Trust grant number 083224 (JPJH) and by the Biotechnology and Biological Sciences Research Council Institute Strategic Programme on Livestock Viral Diseases at The Pirbright Institute and grant BB/H009175/1 (RR).

# Supplementary Information

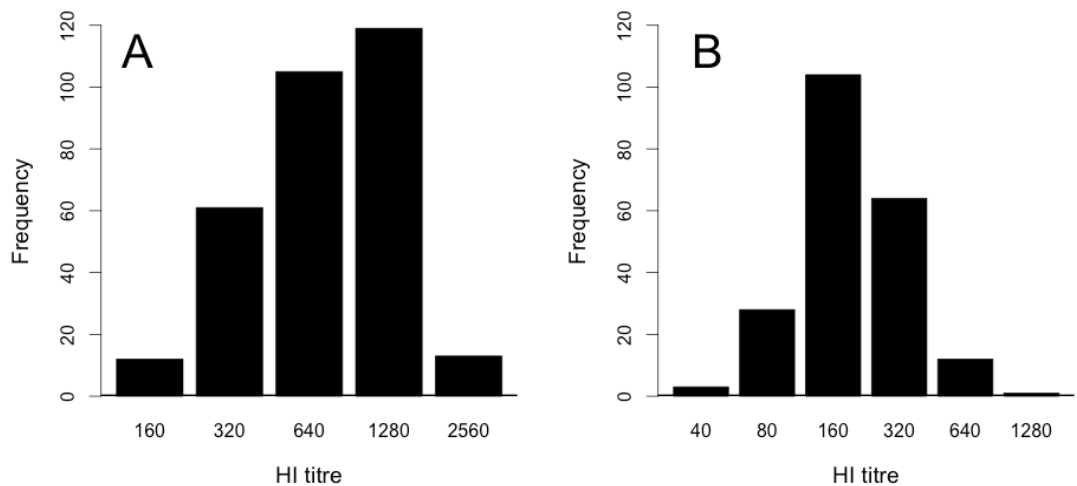

**Figure S1. Distributions of observed HI titers for the most frequently used antigen within the dataset**. Frequency of HI titers recorded for the virus A/New Caledonia/20/99 tested using antisera raised against A/New Caledonia/20/99 (**A: Homologous**) and A/Beijing/262/95 (**B: Heterologous**).

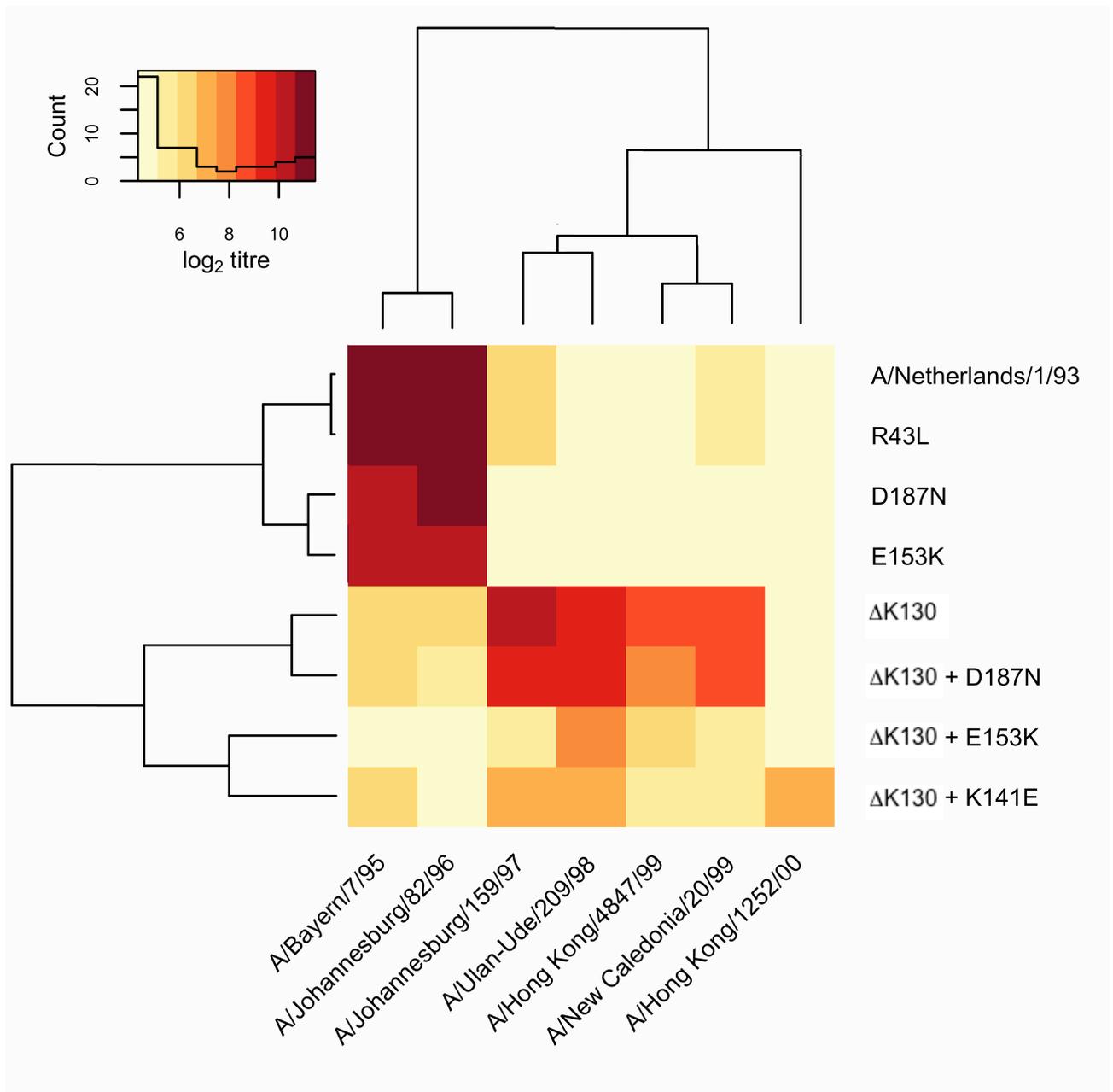

**Figure S2. Heat-map and clustering analysis of HI titers with mutant viruses.** Hierarchical clustering of the wild type Neth93 and mutant viruses generated from it by reverse genetics. Reference strains used to generate antisera are arranged according to their ability to inhibit agglutination of turkey RBCs by each virus. Viruses are simultaneously clustered along the vertical axis according to their antigenic profile. Dendograms indicating antigenic relatedness are shown at the top (for antisera) and to the left (for viruses). Coloring represents $\log_2$ HI titer as indicated at top left with the histogram showing the frequency (count) for each titer.

**Table S1. Antisera used to characterize recombinant viruses**

| Amino acid substitution resulting from mutagenesis | Reference virus against which antisera were raised | | | | | | |
|---|---|---|---|---|---|---|---|
| | A/Bayern /7/95 | A/Johannesburg /82/96 | A/Johannesburg /159/97 | A/Ulan-Ude /209/98 | A/Hong Kong /4847/98 | A/New Caledonia /20/99 | A/Hong Kong /1252/2000 |
| R43L | R | R | L | L | L | L | L |
| ΔK130 | K | K | Δ* | Δ* | Δ* | Δ* | Δ* |
| K141E | K | K | K | K | K | K | E |
| E153K | E | E | E | K | G | G | G |
| D187N | D | D | D | D | N | D | D |

Amino acid identity at HA positions 43, 130, 141, 153 and 187 of reference viruses against which antisera were raised and used to antigenically characterize recombinant viruses (* indicates deletion of amino acid corresponding to position 130). Cells are colored according to whether the reference virus against which antiserum was raised lacked or shared each amino acid substitution introduced by mutagenesis to produce recombinant viruses (R43L, ΔK130, K141E, E153K and D187N). Red indicates that the reference virus lacked the substitution introduced into the recombinant virus and so was in the ancestral state (e.g. R43) and blue indicates that the reference virus shared the introduced substitution (e.g. L43). Absence of color indicates that the reference amino acid identity at the position of substitution in the recombinant virus was different from both of the parental viruses (Neth93 and Neth93 Δ130) and from the mutant virus.

**Table S2. Mean HI titers for recombinant viruses measured against antisera presented in Table S1**

| Antiserum raised against / Parent virus and recombinant viruses | A/Bayern /7/95 | A/Johannesburg /82/96 | A/Johannesburg /159/97 | A/Ulan-Ude /209/98 | A/Hong Kong /4847/98 | A/New Caledonia /20/99 | A/Hong Kong /1252/2000 |
|---|---|---|---|---|---|---|---|
| A/Netherlands/1/93 (Neth93) | 2792 | 2792 | 87 | 28 | 20 | 40 | 20 |
| Neth93 ΔK130 | 87 | 104 | 1174 | 698 | 453 | 453 | 24 |
| Neth93 R43L | 2792 | 2792 | 80 | 28 | 20 | 40 | 20 |
| Neth93 E153K | 1660 | 1522 | 28 | 34 | 20 | 24 | 20 |
| Neth93 D187N | 1396 | 1660 | 698 | 20 | 20 | 28 | 20 |
| Neth93 ΔK130 K141E | 62 | 24 | 174 | 160 | 48 | 48 | 174 |
| Neth93 ΔK130 E153K | 28 | 20 | 57 | 247 | 80 | 44 | 20 |
| Neth93 ΔK130 D187N | 67 | 52 | 698 | 538 | 293 | 320 | 28 |

Geometric mean HI titers are recorded as the reciprocal of the highest dilution of a particular antiserum that inhibited hemagglutination of a standardized concentration of red blood cells by eight hemagglutinating units of each recombinant virus. A visual description of these data is provided in Fig. S2.

**Table S3. Average absolute prediction error (antigenic units) across various predictive models.**

| Antigenic substitutions | Avidity effects | Test error Mean (± SEM) | 95% upper limit on test error |
|---|---|---|---|
| None | None | 1.32 (± 0.0022) | 3.06 |
| Major | None | 1.02 (± 0.0018) | 2.44 |
| Major | Estimated | 0.69 (± 0.0013) | 1.65 |
| All | None | 0.69 (± 0.0012) | 1.65 |
| All | Estimated | 0.54 (± 0.0011) | 1.44 |